\documentclass[draftclsnofoot,12pt,onecolumn,twoside,letterpaper]{IEEEtran}
\usepackage[pdftex]{graphicx}
\usepackage{url}
\usepackage[font = small]{caption}
\usepackage{subcaption}
\usepackage{color}
\usepackage[section]{placeins}
\usepackage[outline]{contour}
\usepackage{cite}
\usepackage{microtype}
\newcommand{\revColor}{}

\begin{document}
%
\title{Full-Duplex Mobile Device -- Pushing\\ the Limits}

\author{Dani~Korpi,
				Joose~Tamminen,
				Matias~Turunen,
				Timo~Huusari,\\
				Yang-Seok~Choi,
				Lauri~Anttila,
				Shilpa~Talwar,
        and~Mikko~Valkama\vspace{-15mm}
\thanks{Manuscript received December 23, 2015; revised April 12, 2016.}%
\thanks{D. Korpi, J. Tamminen, M. Turunen, L. Anttila, and M. Valkama, are with the Department
of Electronics and Communications Engineering, Tampere University of Technology, PO Box 692, FI-33101, Tampere, Finland, e-mail: dani.korpi@tut.fi, joose.tamminen@tut.fi, matias.turunen@tut.fi, lauri.anttila@tut.fi, mikko.e.valkama@tut.fi.}
\thanks{T. Huusari is with Intel Corporation, Tampere, Finland, e-mail: timo.s.huusari@intel.com.}%
\thanks{Y.-S. Choi is with Intel Corporation, Hillsboro, OR 97124, USA, e-mail: yang-seok.choi@intel.com.}%
\thanks{S. Talwar is with Intel Corporation, Santa Clara, CA 95054-1549, USA, e-mail: shilpa.talwar@intel.com.}}%

%



\maketitle


\begin{abstract}
\vspace{-2mm}
In this article, we address the challenges of transmitter-receiver isolation in \emph{mobile full-duplex devices}, building on shared-antenna based transceiver architecture. Firstly, self-adaptive analog RF cancellation circuitry is required, since the capability to track time-varying self-interference coupling characteristics is of utmost importance in mobile devices. In addition, novel adaptive nonlinear DSP methods are also required for final self-interference suppression at digital baseband, since mobile-scale devices typically operate under highly nonlinear low-cost RF components.

In addition to describing above kind of advanced circuit and signal processing solutions, comprehensive RF measurement results from a complete demonstrator implementation are also provided, evidencing beyond 40~dB of active RF cancellation over an 80~MHz waveform bandwidth with a highly nonlinear transmitter power amplifier. Measured examples also demonstrate the good self-healing characteristics of the developed control loop against fast changes in the coupling channel. Furthermore, when complemented with nonlinear digital cancellation processing, the residual self-interference level is pushed down to the noise floor of the demonstration system, despite the harsh nonlinear nature of the self-interference. These findings indicate that deploying the full-duplex principle can indeed be feasible also in mobile devices, and thus be one potential technology in, e.g., 5G and beyond radio systems.
\end{abstract}


\section{Introduction}

\IEEEPARstart{I}{nband} full-duplex communications is widely regarded as one potential solution towards more spectrally efficient wireless networks. The basic idea behind it is to utilize the available temporal and spectral resources to the fullest extent by transmitting and receiving data signals simultaneously at the same center frequency \cite{Sabharwal14,Duarte12}. In theory, this will result in doubling of the radio link data rate while requiring no additional bandwidth. {\revColor{}Furthermore, when combined with proper scheduling in multiuser networks, this can be translated into an increase also in the cell and network capacities \cite{Goyal15a}.} Especially in the future 5G era, inband full-duplex communications can be one enabler and crucial step towards the desired 1000-fold increase in the total throughput \cite{Andrews14,Hong14}. Thus, implementing a fully functional inband full-duplex transceiver is a tempting prospect.

However, in practice, realizing the potential performance gains is far from trivial, as extremely efficient attenuation of the own transmit signal is required. Note that now it is not possible to filter out the own transmission with, e.g., a duplexer, since it is overlapping with the actual received signal of interest in the frequency domain. In theory, canceling this so-called self-interference (SI) can be done by subtracting the own transmit signal from the total received waveform. In practice, on the other hand, the SI signal will always be distorted in a linear as well as nonlinear manner while propagating to the receiver, and thereby it is not a trivial task to reproduce a sufficiently accurate cancellation signal. Attenuating the SI signal by an adequate amount is in fact the central research problem, which must be resolved in order to implement a practical inband full-duplex radio \cite{Sabharwal14,Duarte12,Hong14,Bharadia13}.

\begin{figure*}
        \centering
        \begin{subfigure}[t]{0.45\textwidth}
                \includegraphics[width=\textwidth]{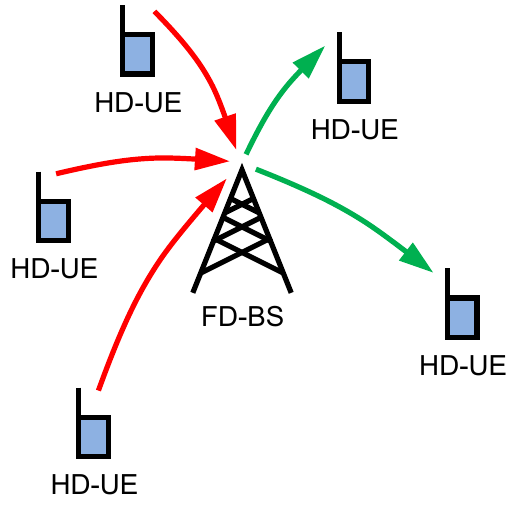}
								\caption{}
                \label{fig:system_hd}
        \end{subfigure}%
        \qquad 
        \begin{subfigure}[t]{0.45\textwidth}
                \includegraphics[width=\textwidth]{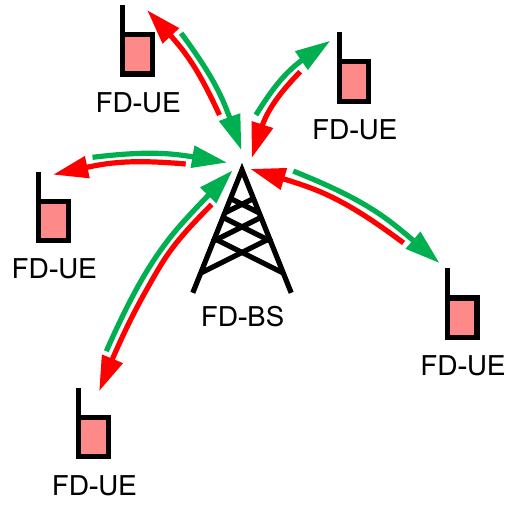}
								\caption{}
                \label{fig:system_fd}
        \end{subfigure}
        \caption{(a) An illustration of a single cell where the base station (BS) is full-duplex capable and the mobiles (UEs) are legacy half-duplex devices, and (b) a similar illustration of a single cell where all parties are full-duplex capable.}
				\label{fig:network}
				\vspace{4mm}
\end{figure*}

The nonlinear distortion due to analog impairments is an especially prevalent issue in mobile-scale devices, which typically utilize low-cost mass-produced RF components. For this reason,  a typical assumption in the reported works has been that, in a mobile cellular network, the base station (BS) is able to communicate in full-duplex mode, whereas the mobiles are legacy half-duplex devices \cite{Everett11}. An illustration of this type of a solution is shown in Fig.~\ref{fig:network}(\subref{fig:system_hd}), where a full-duplex capable BS is serving half-duplex mobile users. The benefit of this solution is that it avoids the challenges of implementing a mobile full-duplex transceiver, and instead requires only the BS to be full-duplex capable. This is a significantly easier prospect, since the BS typically utilizes more expensive higher quality components, and it can also have a significant amount of spatial isolation between the transmitter and the receiver. 

{\revColor{}However, limiting the full-duplex operation only to the BS side does not obviously capitalize the full potential of the full-duplex principle.} By having also the mobile devices full-duplex capable, the overall data-rate of the corresponding cell could be significantly increased \cite{Sahai13}. This is illustrated in Fig.~\ref{fig:network}(\subref{fig:system_fd}), where the BS can now exchange data in a full-duplex manner with each mobile user, resulting in improved spectral efficiency and higher data-rates. For this reason, in this article we will \emph{investigate the possibilities and challenges of implementing full-duplex capable mobile devices}. Due to extreme size and cost constraints in mobile devices, as well as stringent requirements regarding the power consumption, this is an extraordinarily challenging task \cite{Korpi15}.

Firstly, due to the restricted dimensions in a mobile-scale full-duplex transceiver, there is likely to be no space for separate transmit and receive antennas. This means that the transmitter and receiver must share an antenna, while still maintaining a reasonable amount of isolation between each other. There have already been some preliminary demonstrations where this has been implemented in practice, and thereby this aspect of a mobile inband full-duplex transceiver is potentially feasible \cite{Bharadia13,Korpi15}.

{\revColor{}In addition, because of the wide bandwidths of the modern radio systems, advanced wideband cancellation processing in the analog/RF domain is required also in full-duplex mobile devices.} The feasibility of this type of wideband RF cancellation circuits, utilizing, e.g., several delay lines and appropriate amplitude and phase tuning to model and track the frequency and time dependencies of the wideband SI channel, have also been preliminarily demonstrated in practice \cite{Bharadia13,Choi13}. The remaining challenge is implementing a wideband RF canceller in mobile scale, such that the transmitter and receiver utilize the same antenna.

Another aspect, which becomes a considerable factor and concern in mobile devices, is the quality of the analog components. Specifically, the low-cost components, which are typically used in mass-produced handheld devices, distort the SI signal such that linear digital processing alone cannot reproduce and cancel the residual SI waveform accurately enough \cite{Korpi15,Bharadia13}. This means that advanced modeling and processing, taking into account the different analog impairments, is required in order to produce a sufficiently accurate cancellation signal. For instance, modeling nonlinear distortion in the digital SI regeneration and cancellation stage has been shown to improve the performance of a practical inband full-duplex transceiver \cite{Bharadia13,Korpi15}.


{\revColor{}In this article we will take a closer look into the aforementioned challenges. In addition, we will also present some of our recent findings for solving them and show with an actual prototype implementation that the challenges caused by the limited size and RF component quality of mobile-scale devices can be tackled by incorporating state-of-the-art algorithms and cancellation processing. In particular, this article builds partially on the recent scientific findings by the authors, reported primarily in \cite{Choi13,Korpi13,Korpi133,Korpi15}.}

The rest of this article is organized as follows. In Section~\ref{sec:circ_architecture}, we will review the general architecture of a shared-antenna full-duplex mobile transceiver, alongside with the challenges faced by such a device. Then, in Section~\ref{sec:rf_canc}, a self-adaptive multi-tap RF cancellation architecture is presented, enabling efficient cancellation and tracking of wideband SI signals already at the RF stage. After this, in Section~\ref{sec:nl_canc}, we will discuss state-of-the-art digital algorithms for modeling the nonlinear distortion produced in the transmitter chain, most notably in the power amplifier, and how to incorporate this in the final digital SI cancellation stage. Then, in Section~\ref{sec:results}, a prototype implementation of a mobile full-duplex transceiver is evaluated with RF measurements, which show that the considered architecture can cancel the self-interference to the level of the receiver noise floor. Finally, the conclusions are drawn in Section~\ref{sec:conc}.

\section{Shared-Antenna Mobile Full-Duplex Device Architecture}
\label{sec:circ_architecture}

\begin{figure*}[!t]
\centering
\begin{subfigure}[t]{\textwidth}
\centering
\includegraphics[width=\textwidth]{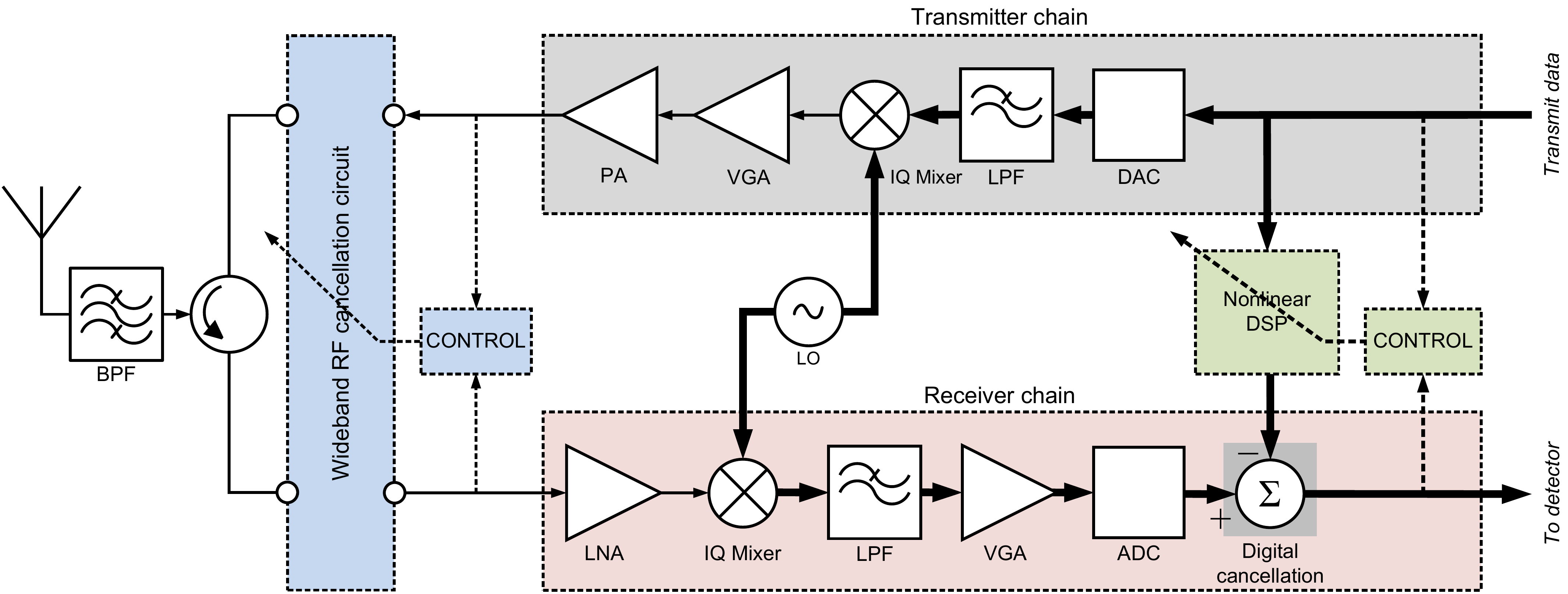}
\caption{}
\label{fig:block_diagram}
\vspace{4mm}
\end{subfigure}
\begin{subfigure}[b]{0.7\textwidth}
\centering
\includegraphics[width=\textwidth]{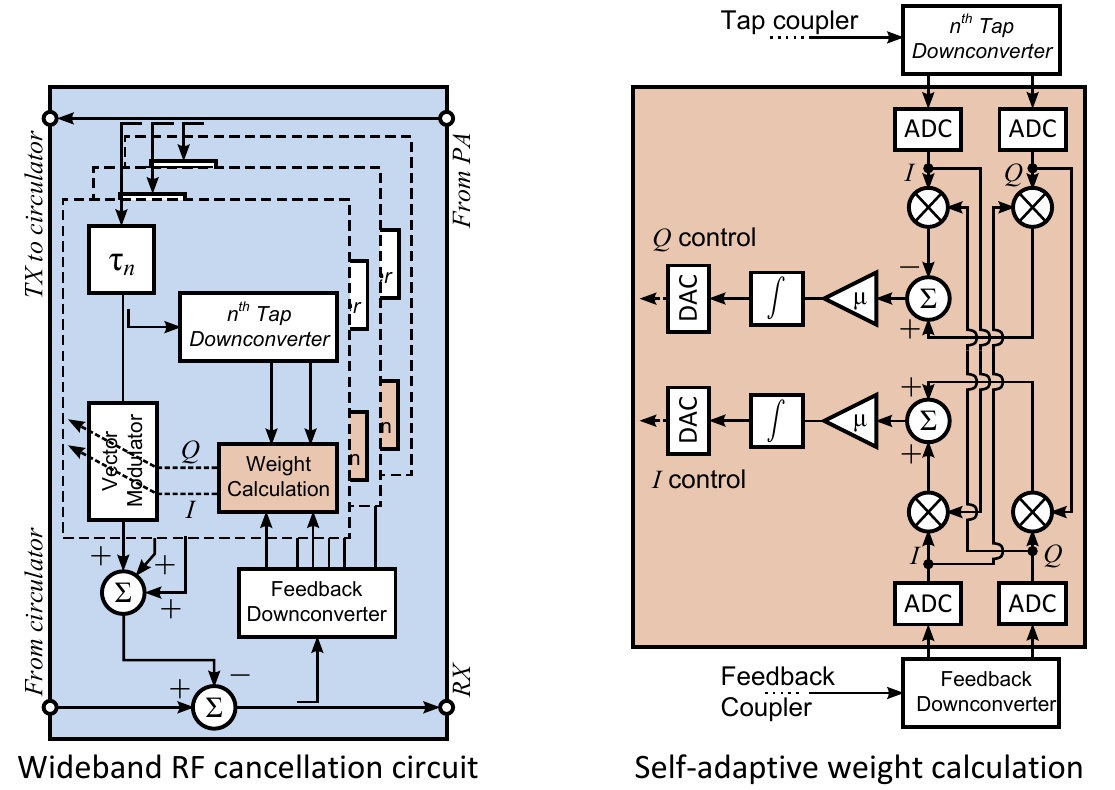}
\caption{}
\label{fig:rf_cancel_block}
\end{subfigure}
\caption{(a) The considered shared-antenna mobile full-duplex transceiver architecture, including self-adaptive RF cancellation and self-adaptive nonlinear digital cancellation, alongside with (b) a general block diagram depicting the RF canceller and the self-adaptive weight calculation for one tap.}
\label{fig:system}
\vspace{-5mm}
\end{figure*}

In this article, we consider the challenges faced by a shared-antenna mobile-scale direct-conversion full-duplex transceiver. The general structure of such an inband full-duplex device is shown in Fig.~\ref{fig:system}(\subref{fig:block_diagram}), which illustrates all the relevant aspects required to achieve real full-duplex operation under practical conditions. The direct-conversion radio architecture is a natural choice for the considered full-duplex transceiver as it is the most widely used structure for modern wireless radios, especially in mobile-scale devices.


The key component enabling the transmitter and receiver to share a single antenna in a mobile full-duplex device is a circulator, which is used to connect the antenna to the transceiver. {\revColor{}A circulator is a three port device which steers the signal through its ports such that it comes in at one port and then exits the circulator from the next port, depending on the direction of rotation (i.e., clockwise or counterclockwise). In principle, the signal cannot propagate to the opposite direction, which ensures a certain amount of isolation between the transmitter and the receiver. Depending on the size and cost of the circulator, typical practical values for the isolation vary between 20 and 60~dB, while the attenuation in the desired direction is usually less than half a decibel.} Circulators are generally bandwidth limited devices such that wideband operation can be attained at the expense of worse overall isolation. Being a passive component, its size is ultimately dictated by the wavelength of the operating frequency.

{\revColor{}Another option for isolating the transmitter and receiver sharing the same antenna is to use an electrical balance duplexer, which essentially tries to mimic the impedance of the antenna with a tunable balance network \cite{Debaillie14}. With the help of a hybrid transformer, this results in a significant amount of isolation between the transmitter and the receiver. An electrical balance duplexer can be implemented in a more compact form than a circulator, which makes it very suitable for mobile devices, but it also suffers from some inherent insertion loss. Another drawback of the electrical balance duplexer is the need to actively tune the impedance network, since the impedance of the antenna is time-variant. This is obviously not needed in a circulator-based system. For these reasons, in this article we focus on a demonstrator implementation utilizing a circulator, since it is a passive device and thereby more robust. Nevertheless, for future implementations, the electrical balance duplexer is also a very prominent candidate.}

When using the circulator-based architecture, there are two strong components in the SI signal, observed towards the receiver path. Firstly, there will be leakage through the circulator, whose magnitude can be estimated by subtracting the amount of circulator isolation from the transmit power. Here, the SI is usually attenuated by at least 20~dB, as already mentioned. The second strong component is the power reflected by the mobile antenna, caused by the impedance mismatch at its input. If the input was perfectly matched, the antenna would accept all the supplied power and this SI component would not exist, but in practice the mismatch will always cause a part of the power to be reflected back to the transmission line. Typical target value for antenna matching in mobile devices is around $-10$~dB as better matching does not significantly increase the transmitted power. This obviously results in a very powerful reflected component in an inband full-duplex radio, and thereby higher values of matching are desirable as they translate directly to a reduced amount of SI. Nevertheless, when using off-the-shelf antennas, matching values better than $-20$~dB are seldom obtained, and the reflection from the antenna constitutes a significant portion of the total SI, potentially even dominating compared to direct leakage through the circulator.

Weaker components in the composite SI signal come mainly from the multipath reflections, which propagate back to the antenna from the surrounding environment. They are heavily dependent on the type of environment around the antenna but usually the multipath reflections will be significantly weaker than the leakage through the circulator or the reflection from the antenna. However, a change in the near field of the antenna (e.g., wrapping a hand around it) affects its matching, which will directly change the amount of reflected power.

Because of the leakage through the circulator, as well as the reflections coming from the antenna and the surrounding environment, additional SI attenuation is typically required, both in the analog/RF and digital domains. In general, the overall analog attenuation of the SI signal prior to entering the receiver chain must be sufficient to ensure that
\begin{itemize}
\item The SI power level is not too high for the receiver low-noise amplifier (LNA), to prevent receiver saturation
\item The dynamic range of the analog-to-digital-converters (ADCs) is high enough to capture the residual SI as well as the weak received signal of interest with sufficient precision.
\end{itemize}
Depending on the receiver, either of these can be the limiting factor \cite{Korpi13}. Usually, the passive SI attenuation provided by circulator isolation and antenna matching is clearly insufficient to ensure these requirements \cite{Choi13,Korpi13}. This creates a strong motivation for active RF cancellation, which provides additional SI suppression before the actual receiver chain by subtracting a modified copy of the transmit signal from the overall received signal.

Remembering again the wide bandwidth of the signals that are used in modern cellular networks, it is obvious that the active RF canceller within a mobile full-duplex device must be capable of efficient \emph{wideband cancellation}. This can be ensured by having a multi-tap analog SI canceller where several differently delayed copies of the transmit signal are used as reference signals, each of them having tunable amplitude and phase. This type of an RF canceller is capable of modeling the coupling channel over significantly wider bandwidths than the conventional solutions \cite{Bharadia13,Choi13}. The objective of the RF cancellation circuit is to match the phases and amplitudes of the reference signals such that the produced cancellation signal matches with the composite SI signal, coming from the circulator and the antenna, at the summing node just before the receiver chain. {\revColor{}However, the cancellation signal needs to be adjusted to be 180 degrees out of phase to ensure destructive superposition and hence obtain cancellation. Also note that, by using the transmitter output as a reference signal, all of the transmitter-induced impairments are implicitly taken into account by the RF canceller \cite{Choi13,Korpi13}.}

Another important consideration for a mobile full-duplex device is to also have sufficient adaptivity in the RF canceller. Namely, to support efficient cancellation of the SI under practical conditions, the control of the phases and amplitudes of the cancellation signals needs to be \emph{self-adaptive} in order to track sudden changes in the close proximity of the antenna. {\revColor{}These changes in the SI channel environment are caused by the moving objects in the vicinity of the device, such as a person walking by, or by the movement of the device itself. By monitoring the signal power level at the canceller output, the control can be made automatic by using either digital or analog tracking circuits \cite{Choi13}.} This self-tuning of the analog SI cancellation signal is perhaps one of the most crucial elements of a practical mobile inband full-duplex radio, and is typically neglected in most of the reported works in this field so far \cite{Duarte12}. This topic, alongside with other aspects related to the RF canceller, is elaborated in more details in Section~\ref{sec:rf_canc}.

{\revColor{}Furthermore, due to the high power level of the received SI signal, analog SI cancellation alone is typically not enough to attenuate it below the receiver noise floor.} Thus, the final attenuation of the residual SI must be done in the \emph{digital domain}. There, the cancellation signal can be constructed from the original transmit data by filtering it in accordance with the remaining effective SI channel. One important benefit of digital SI cancellation is the relatively easy inclusion of nonlinear modeling of the SI waveform, which can be done conveniently by utilizing nonlinear basis functions \cite{Bharadia13,Korpi15}, as well as the natural support for self-tracking of the SI channel characteristics through adaptive filtering. As has been demonstrated recently, nonlinear modeling can significantly improve the SI cancellation performance in a practical full-duplex transceiver \cite{Bharadia13,Korpi15}. Thus, nonlinear adaptive digital signal processing is also a key feature in a mobile full-duplex device to ensure efficient cancellation and tracking of residual SI. This is addressed in more details in Section~\ref{sec:nl_canc}.

\section{Advanced Self-Adaptive RF Cancellation Principle}
\label{sec:rf_canc}

This section describes the detailed principle of such an RF cancellation circuit that fulfills the aforementioned requirements regarding wideband operation and self-adaptivity \cite{Choi13}. A prototype implementation of this type of an RF canceller is then reported and measured in Section~\ref{sec:results}. As outlined above, the RF canceller aims at reconstructing and canceling the received composite SI waveform, which consists of various components with different delays. Since the delays of these components are unknown and also time varying, the delays in the SI regeneration and cancellation paths are not equal to the true delays. Hence, as a whole, the active RF canceller can be seen as an interpolator which tries to regenerate and track the true composite SI, by using \emph{predefined delays but tunable amplitudes and phases} \cite{Choi13}.

In general, optimum linear filtering for interference or echo cancellation is a thoroughly studied field in the literature. However, most of the reported research and implementations focus on digital baseband while here our focus is fully at analog RF domain. In the full-duplex radio field, optimum filtering based analog RF cancellation has been recently addressed, in terms of passband analog finite impulse response (FIR) filtering, in \cite{Bharadia13} and \cite{Choi13}. The filter coefficients or weights can be obtained in analog \cite{Choi13} or digital domain \cite{Bharadia13}, and in adaptive \cite{Choi13} or non-adaptive \cite{Bharadia13} manner. Unlike in \cite{Bharadia13}, \emph{per-tap phase shifting} is also included in \cite{Choi13}, allowing phase rotation of a tapped delayed signal at passband. Thus, the filter tap weights in \cite{Choi13} become complex, when interpreted from the baseband waveform perspective, while those in \cite{Bharadia13} are real. These effectively complex taps significantly reduce the cancellation performance dependency on frequency, tap delays, and the underlying true delays of the different SI components, thereby also reducing the number of taps required. This, in turn, is crucial for mobile devices in order to minimize cost, size, and power consumption.

In general, there are two options for obtaining and controlling the tap weights: an \emph{open loop} and a \emph{closed loop}. In the \emph{open loop}, separate SI channel estimation is needed, followed by the actual canceller weight calculations. Such strategy obviously calls for digital processing of a large amount of data, thereby producing a significant delay in the canceller adaptation. In the \emph{closed loop}, on the other hand, the weights are directly optimized to minimize the SI power at the canceller output. Such closed-loop adaptive processing structure is thus essentially a negative feedback system, where the weights are automatically adjusted in real-time to keep the residual SI power low at the canceller output. This strategy is very well suitable for directly tracking a time-varying SI channel under strict delay requirements \cite{Choi13}.


For this reason, in a mobile device, the weight adaptation for RF cancellation must be done in a closed-loop fashion since tracking the characteristics of the overall SI waveform in real-time is a crucial feature. The general structure for such a closed-loop wideband RF canceller circuit utilizing three taps is illustrated in Fig.~\ref{fig:system}(\subref{fig:rf_cancel_block}), where the LMS-based learning algorithm is also shown for a single tap. This type of a canceller structure has been observed to provide excellent cancellation performance under a wide bandwidth and highly varying channel conditions, as will be shown through measurements in Section~\ref{sec:results}. Furthermore, as discussed in detail in \cite{Choi13}, this type of an RF canceller is very robust against various circuit imperfections that typically occur in mobile-scale devices. In particular, deploying the power amplifier (PA) output as the reference signal in RF cancellation is beneficial, since this way all the main transmit chain imperfections are automatically included in the cancellation signal \cite{Choi13,Korpi13}, and hence subtracted along with the linear SI. This is particularly important when utilizing a mobile-scale power-efficient PA, which creates substantial nonlinear distortion. Thus, an analog RF canceller, with PA output as the reference signal, provides significant immunity to transmitter impairments, including nonlinear distortion, phase noise, and even transmitter noise \cite{Choi13}. This then relaxes to certain extent the requirements on the final residual SI suppression at digital baseband, and also reduces the required dynamic range for the main receiver ADC.

\section{Adaptive Nonlinear Digital Cancellation for Final SI Suppression}
\label{sec:nl_canc}

After analog SI cancellation, the power level of the residual SI can still be relatively strong in the digitized signal. This calls for additional digital SI cancellation, which will then decrease the level of the SI signal below the receiver noise floor. The most straight-forward method for canceling SI in the digital domain is to use the original transmit data as the reference signal, which is then filtered according to the effective channel experienced by the residual SI signal and subsequently subtracted from the overall received signal. The channel includes the effects of both the transmitter and the receiver, the circulator, and the RF canceller, as well as the multipath components reflected from the antenna and the surrounding environment. Modeling, estimating and tracking this effective SI channel is the key factor in digital SI cancellation, and it will determine the achievable cancellation in the digital domain.

Typically, in most works reported in the literature, the effective SI channel is assumed to be a linear multipath channel, which essentially means that the transmitter and receiver chains are assumed to be linear \cite{Duarte12}. With high-quality components, e.g., well-calibrated laboratory equipment, this can indeed be the case. Then, to perform efficient digital SI cancellation, it is sufficient to obtain a linear channel estimate between the original transmit data and the observed SI signal. However, when considering a mobile-scale full-duplex transceiver utilizing low-cost mass-produced components, assuming the transmitter and receiver chains to be linear will result in a significant model mismatch. {\revColor{}In particular, the transmitter PA is typically heavily nonlinear, especially with the higher transmit powers.} This has a significant impact on the residual SI observed at digital baseband \cite{Korpi13,Korpi15}.

\begin{figure*}[!t]
\centering
\includegraphics[width=0.7\columnwidth]{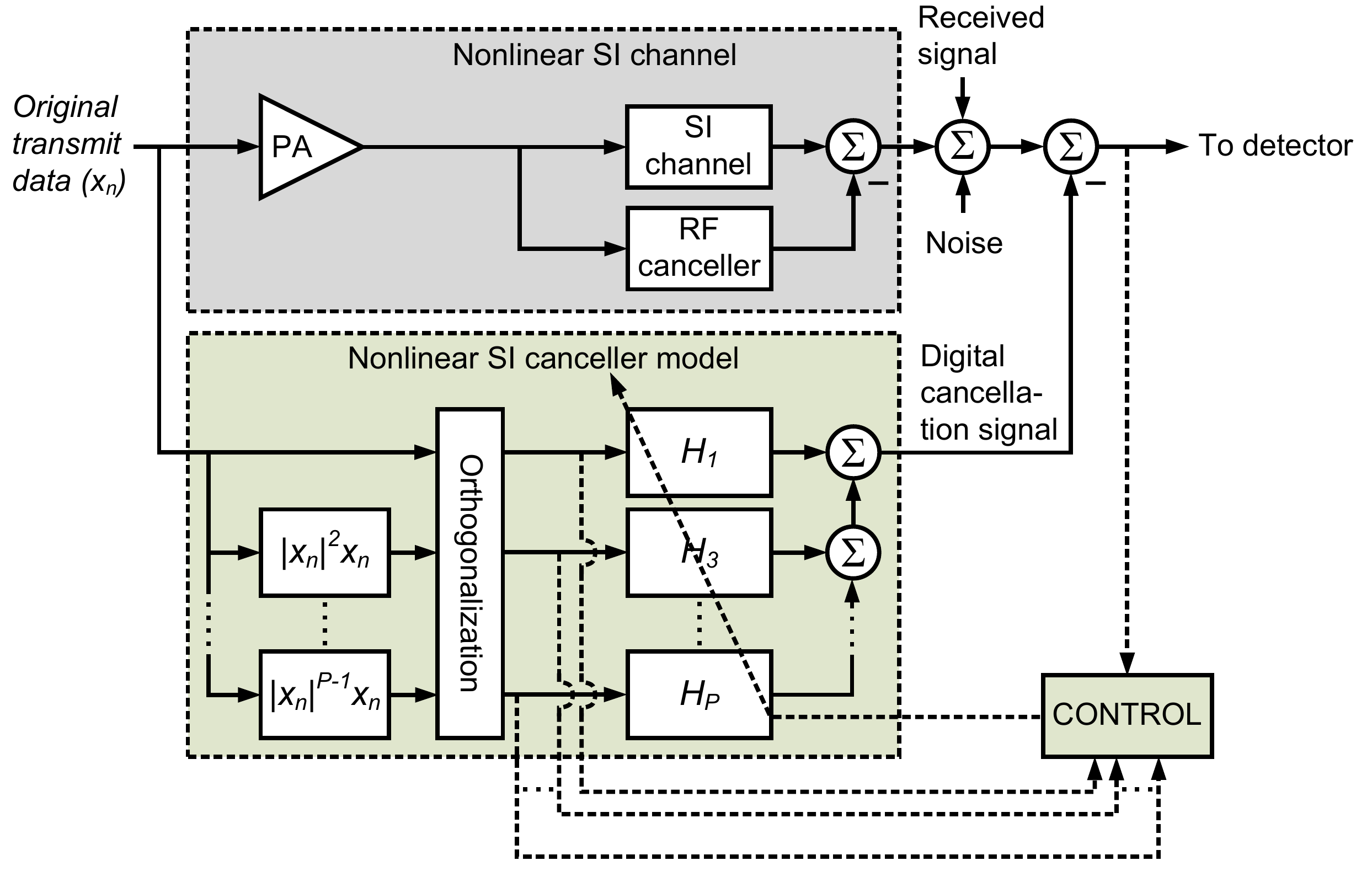}
\caption{A general illustration of adaptive nonlinear digital cancellation processing, incorporating behavioral models for a wideband nonlinear power amplifier and time-varying coupling characteristics.}
\label{fig:nl_canceller}
\vspace{-9mm}
\end{figure*}

{\revColor{}Stemming from the above, the nonlinearity of the components must be considered in the digital cancellation processing.} Especially, the nonlinear distortion produced by the transmitter PA significantly alters the waveform of the SI signal, and thereby it must be included in the SI channel model \cite{Bharadia13,Korpi15}. In principle, this can be done by modeling the residual SI as a weighted sum of nonlinear transformations of the original transmit data, each of which has also some delayed components (memory) present. A principal structure of such a nonlinear digital canceller is shown in Fig. \ref{fig:nl_canceller}. In this nonlinear canceller, the actual transceiver chain is modeled as a cascade of a nonlinear PA and a linear filter, the latter of which consists of the PA memory, multipath components of the SI signal reflected from the surroundings, and the RF cancellation circuit \cite{Korpi15}. This means that the nonlinear residual SI channel follows a parallel Hammerstein (PH) model, whose parameters are relatively straight-forward to estimate and track \cite{Isaksson06}.

{\revColor{}The actual estimation and cancellation procedure involves first transforming the original transmit data with \emph{nonlinear basis functions}, and then orthogonalizing these transformations in order to ensure efficient learning, as discussed in more detail in \cite{Korpi15}. After this, the filter coefficients for each orthogonalized nonlinear transformation are adaptively estimated based on the observed SI signal \cite{Korpi15}.} These estimated coefficients form a memory model for each nonlinearity order, in essence meaning that now the SI channel is estimated and tracked separately for each nonlinear transformation of the transmit data, instead of just the original transmit signal.

In general, the parameter estimation can be carried out, for instance, with block least squares or least mean squares (LMS), depending on the application and available computational resources. In a practical mobile transceiver, adaptivity is a very important factor, as already discussed, and thus LMS or some other adaptive algorithm is preferred to ensure high performance under varying coupling channel conditions. In Figs.~\ref{fig:system}(\subref{fig:block_diagram}) and~\ref{fig:nl_canceller}, the adaptivity is depicted by the real-time control block, which tunes the coefficients based on the canceller output signal. The digital canceller output signal is also used for the actual receiver digital baseband processing, including the detection of the actual received signal of interest.

The performance of this type of a nonlinear digital canceller is of course highly dependent on the validity of the underlying model. Although the assumed PH model has been shown to be quite accurate for a wide variety of PAs \cite{Isaksson06}, in practice it is not necessarily able to achieve perfect accuracy in modeling an arbitrary PA. For example, any frequency selectivity before the PA input will result in a model mismatch. In addition, there are other sources of impairments that are obviously not included in the PH model, such as phase noise \cite{Syrjala13}, which introduces additional noise into the system and thereby decreases the accuracy of the estimated coefficients. However, under typical circumstances, the nonlinearity of the PA is the most significant analog impairment from the SI cancellation perspective \cite{Korpi13}, which means that the PH model can be expected to provide sufficient SI cancellation performance.

Overall, even though a nonlinear cancellation algorithm will obviously require additional resources due to increased computational complexity, modeling the PA-induced nonlinearities adaptively is necessary in order to obtain sufficient levels of digital SI cancellation in a mobile full-duplex transceiver \cite{Korpi13}. Combining an adaptive nonlinear digital canceller with the proposed multi-tap adaptive RF canceller will result in an agile mobile full-duplex transceiver design that is capable of conforming to the surrounding channel environment in both the analog and digital domains. In the following, we will demonstrate this with an actual implemented inband full-duplex transceiver prototype by integrating the different considered cancellation stages together, while deploying a low-cost mass-produced PA driven deep to nonlinear operation.

\section{Demonstrator Implementation and Measured Results}
\label{sec:results}

\begin{figure*}[!t]
\centering
\includegraphics[width=0.5\columnwidth]{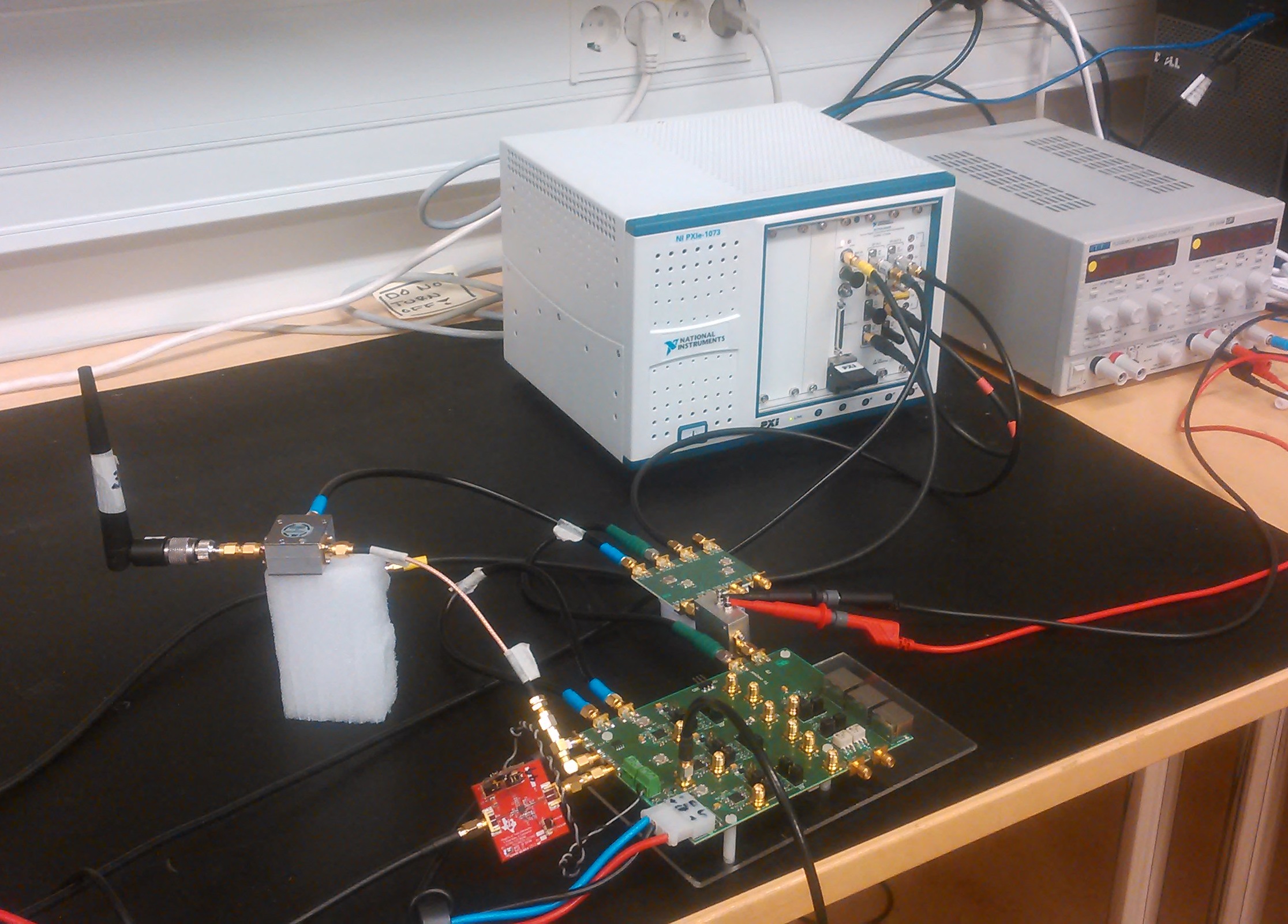}
\caption{The laboratory measurement setup used to evaluate and demonstrate the analog and digital self-interference cancellation performance of a shared-antenna full-duplex transceiver.}
\label{fig:meas_setup}
\end{figure*}

{\revColor{}To evaluate the total performance of the described mobile full-duplex transceiver architecture, real-life RF measurements and experiments are performed with the measurement setup shown in Fig.~\ref{fig:meas_setup}. The measurements are carried out using a National Instruments PXIe-5645R vector signal transceiver (VST) both as a transmitter and a receiver, complemented with an external PA. The used transmit signal is an LTE waveform with an instantaneous bandwidth of 20, 40 or 80 MHz, centered at 2.46~GHz. The VST output is then connected directly to a Texas Instruments CC2595 PA which has a gain of 24~dB at the chosen input power level. The used PA is a commercial low-cost chip intended to be used in low-cost battery-powered devices. This means that the PA produces a significant amount of nonlinear distortion into the SI waveform, especially with the input power levels used in these measurements.}

After the PA, the signal is divided between the RF canceller prototype interface and the antenna port using a directional coupler. Accounting for all the losses incurred by dividing the transmit signal among the different paths, the approximate transmit power at the antenna is in the order of +6$\dots$+8~dBm, depending on the bandwidth. {\revColor{}Such transmit powers are feasible in future ultra-dense 5G networks where inter-site distances below 100~m are to be expected\footnote{Nokia Solutions and Networks, "Ten key rules of 5G deployment," \textit{white paper C401-01178-WP-201503-1-EN}, 2015}, while experimenting with higher transmit powers is an important topic for our future work.} The deployed circulator and the low-cost shared-antenna yield an overall isolation only in the order of 20~dB between the transmitter and the receiver chains, mostly because of the reflection from the antenna. {\revColor{}Then, after the circulator, the desired RX signal and SI are routed back to the prototype RF canceller, which performs the RF cancellation procedure utilizing the PA output signal as described in Section~\ref{sec:rf_canc}.} Finally, the RF cancelled signal is routed to the receiver (NI PXIe-5645R) and captured as digital I and Q samples, which are post-processed offline to implement linear as well as nonlinear digital baseband cancellation. The parameter learning in the digital canceller is done with basic LMS adaptation described in more detail in \cite{Korpi15}, using a highest nonlinearity order ($P$) of 11. {\revColor{}Learning the parameters with such a simple algorithm guarantees that the digital cancellation procedure is done in a computationally efficient manner, which is obviously an important aspect in a mobile-scale device. In the forthcoming results, the adaptive digital cancellation algorithm is first allowed to converge towards the steady-state coefficient values, after which the cancellation performance is measured.} This ensures that the results show the true performance of the digital canceller.

\subsection{Self-Adaptive RF Canceller Implementation and Measured Performance}

\begin{figure*}
        \centering
        \begin{subfigure}[t]{0.32\textwidth}
                \includegraphics[width=\textwidth]{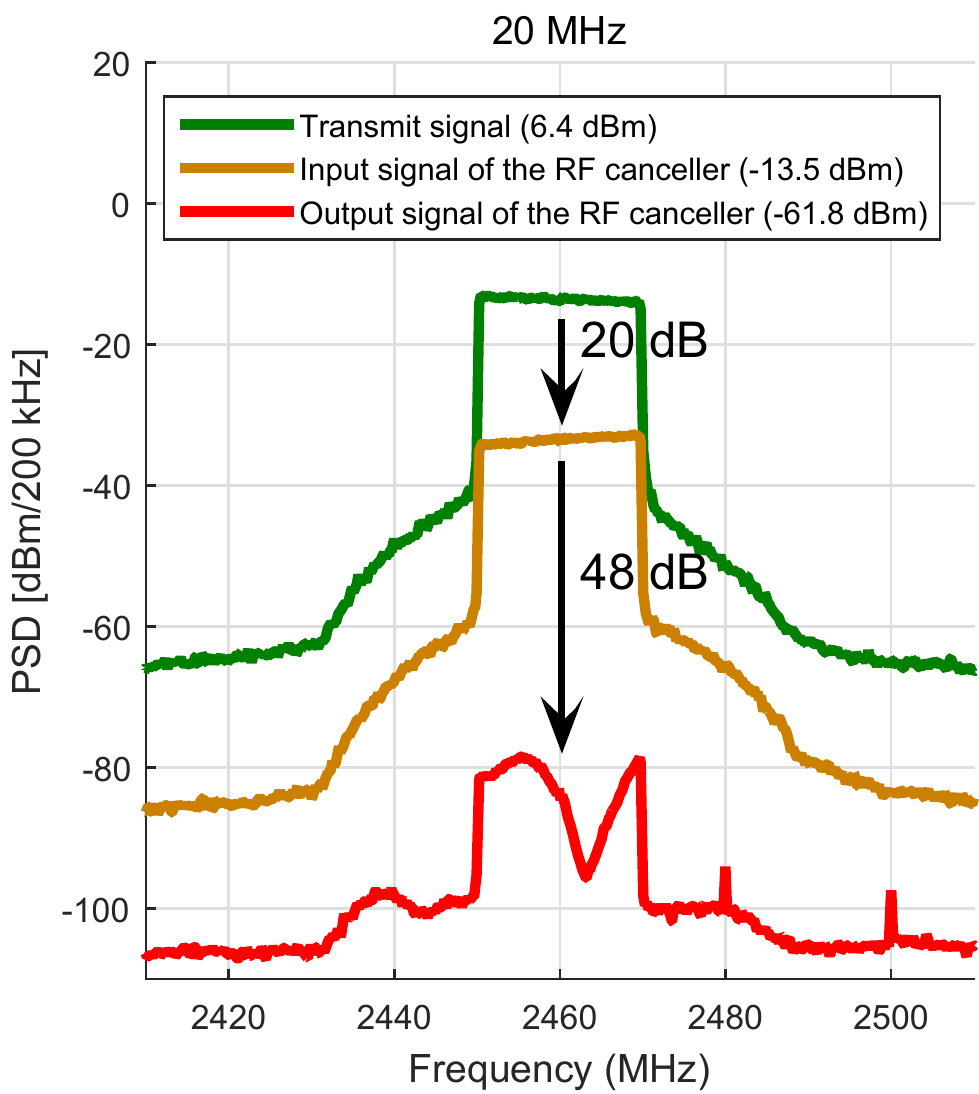}	
								\caption{}
                \label{fig:rf_cancel_20M}
        \end{subfigure}%
        ~
        \begin{subfigure}[t]{0.32\textwidth}
                \includegraphics[width=\textwidth]{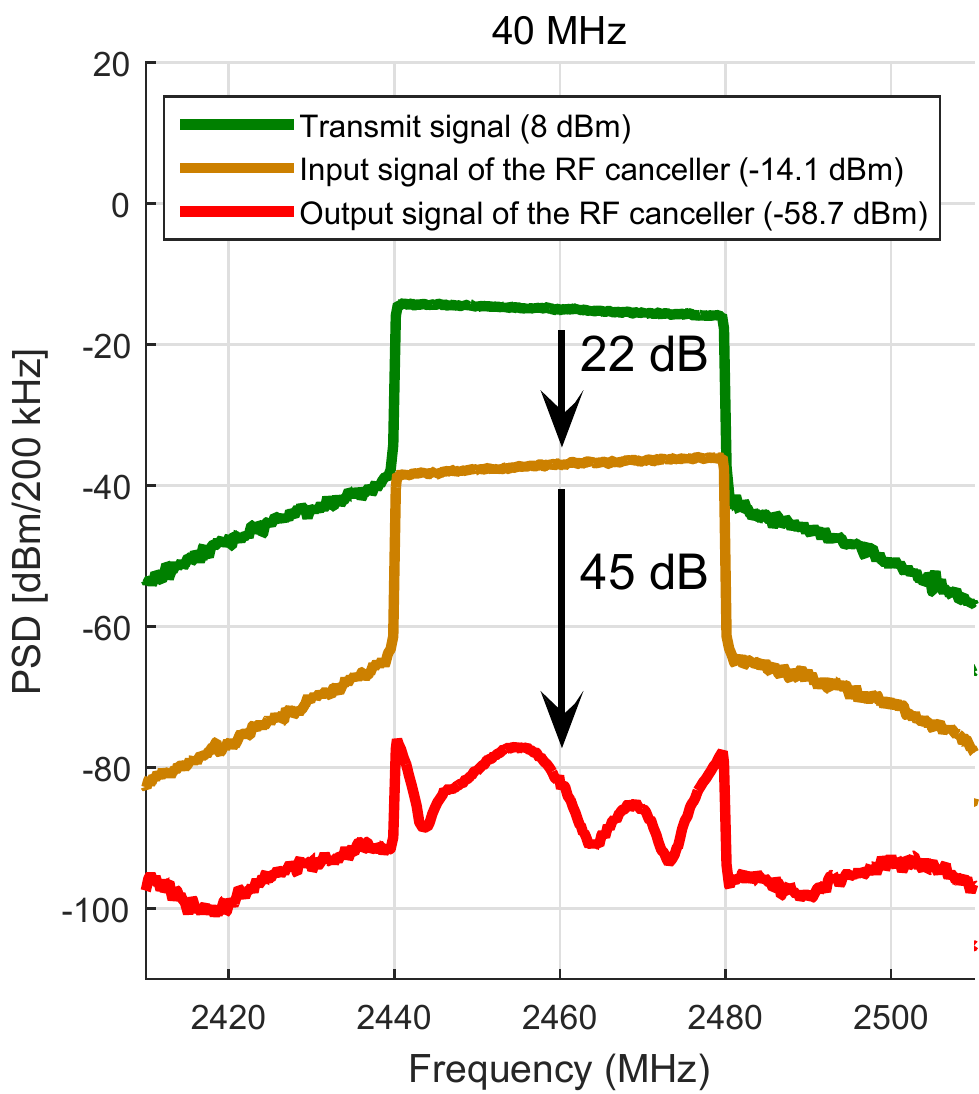}
								\caption{}
                \label{fig:rf_cancel_40M}
        \end{subfigure}
				~
        \begin{subfigure}[t]{0.32\textwidth}
                \includegraphics[width=\textwidth]{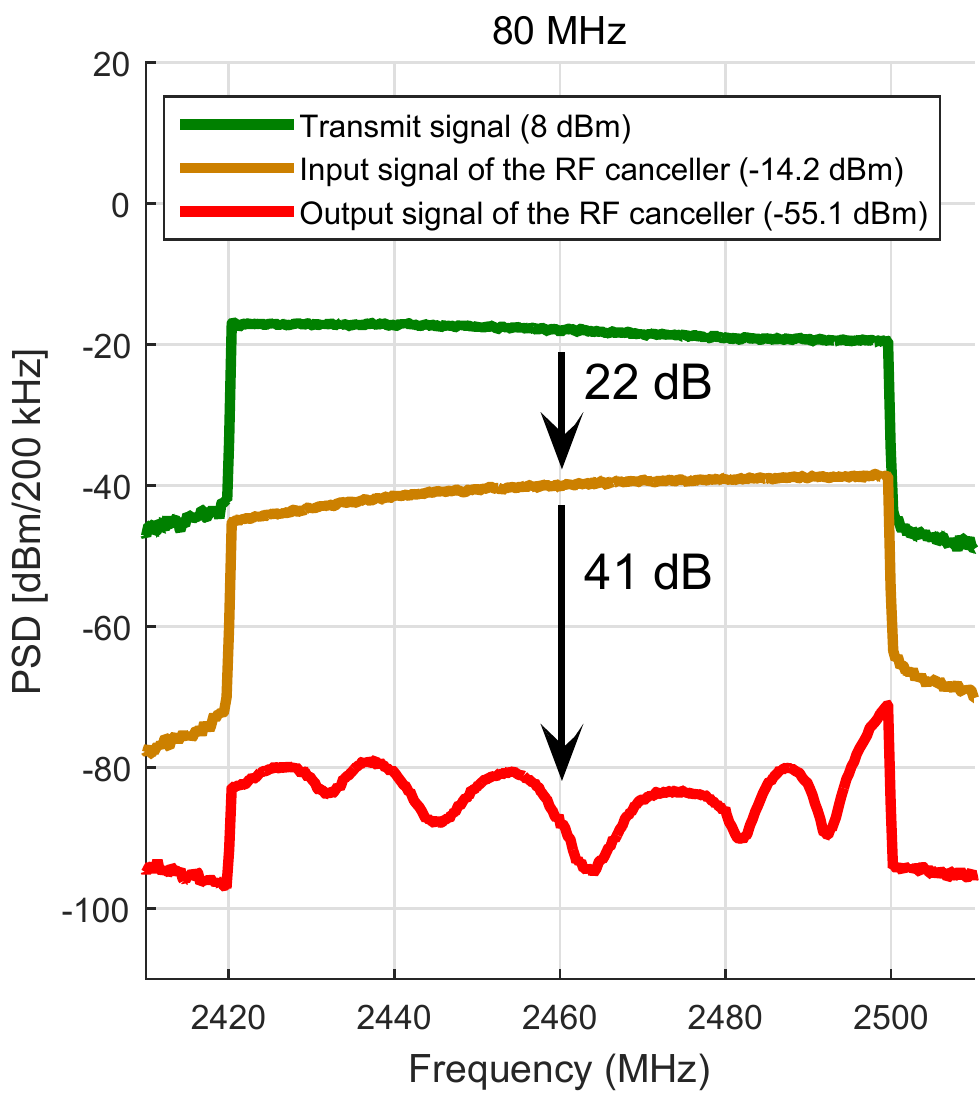}
								\caption{}
                \label{fig:rf_cancel_80M}
        \end{subfigure}
        \caption{The RF cancellation performance at 2.46~GHz center-frequency with (a) 20~MHz, (b) 40~MHz, and (c) 80~MHz transmit signals. The value in the parentheses is the power measured over the mentioned bandwidth around the center frequency.}\label{fig:rf_cancel}
\end{figure*}

First, the performance of the self-adaptive RF canceller, elaborated in more details in Section~\ref{sec:rf_canc}, is evaluated with different bandwidths. The implemented RF cancellation circuit utilizes three taps, their respective controls being implemented by using digital baseband signals and LMS learning as illustrated in Fig.~\ref{fig:system}(\subref{fig:rf_cancel_block}). The amplitude and phase control, per path, is performed using an analog vector modulator, whose I and Q control voltages are given by the self-adaptive control block presented in Fig.~\ref{fig:system}(\subref{fig:rf_cancel_block}). In the control block, the RF signals are first downconverted and digitized, after which they are used for LMS learning. The vector modulators are then provided with the control signals given by the LMS algorithm. The cancellation signal is formed by summing up the phase- and amplitude-adjusted RF signals, after which it is subtracted from the overall received signal. The cancelled signal is then routed both to the actual receiver chain and to the RF cancellation control block to be used in the next LMS iteration.

Figures~\ref{fig:rf_cancel}(\subref{fig:rf_cancel_20M}),~\ref{fig:rf_cancel}(\subref{fig:rf_cancel_40M}), and~\ref{fig:rf_cancel}(\subref{fig:rf_cancel_80M}) show the measured RF signal spectra after the RF canceller when the signal bandwidth is 20, 40, and 80~MHz, respectively. Also the spectra of the transmit signal and the RF canceller input signal are shown for reference. With the 20~MHz signal, the overall suppression achieved by the RF canceller is close to 50~dB, and even with the 80~MHz signal the SI can still be attenuated by more than 40~dB. To the best of our knowledge, these are the highest reported values for \emph{active RF cancellation} thus far, especially for such wideband signals. By increasing the number of taps in the canceller, even wider bandwidths can be potentially supported with this architecture \cite{Choi13}.

Next, we demonstrate the self-adaptation capabilities of the developed RF cancellation circuit and its underlying automated control intelligence. A time-varying reflection scenario around the antenna is deliberately created by bringing different reflecting materials close to the antenna. This changes heavily the total SI coupling channel, and hence calls for fast adaption in the RF cancellation circuit. This overall setup is demonstrated through a video recording of the experimentation, showing that the developed RF canceller can rapidly self-heal its operation by automatically tuning the amplitudes and phases of the RF cancellation paths. The video recording is available at
\begin{itemize}
\item RF canceller self-healing demonstration: {\color{blue}\url{http://www.tut.fi/full-duplex/commag}}
\end{itemize}

\subsection{Total Integrated System Performance Including Nonlinear Digital Cancellation}

\begin{figure*}
        \centering
        \begin{subfigure}[t]{0.32\textwidth}
                \includegraphics[width=\textwidth]{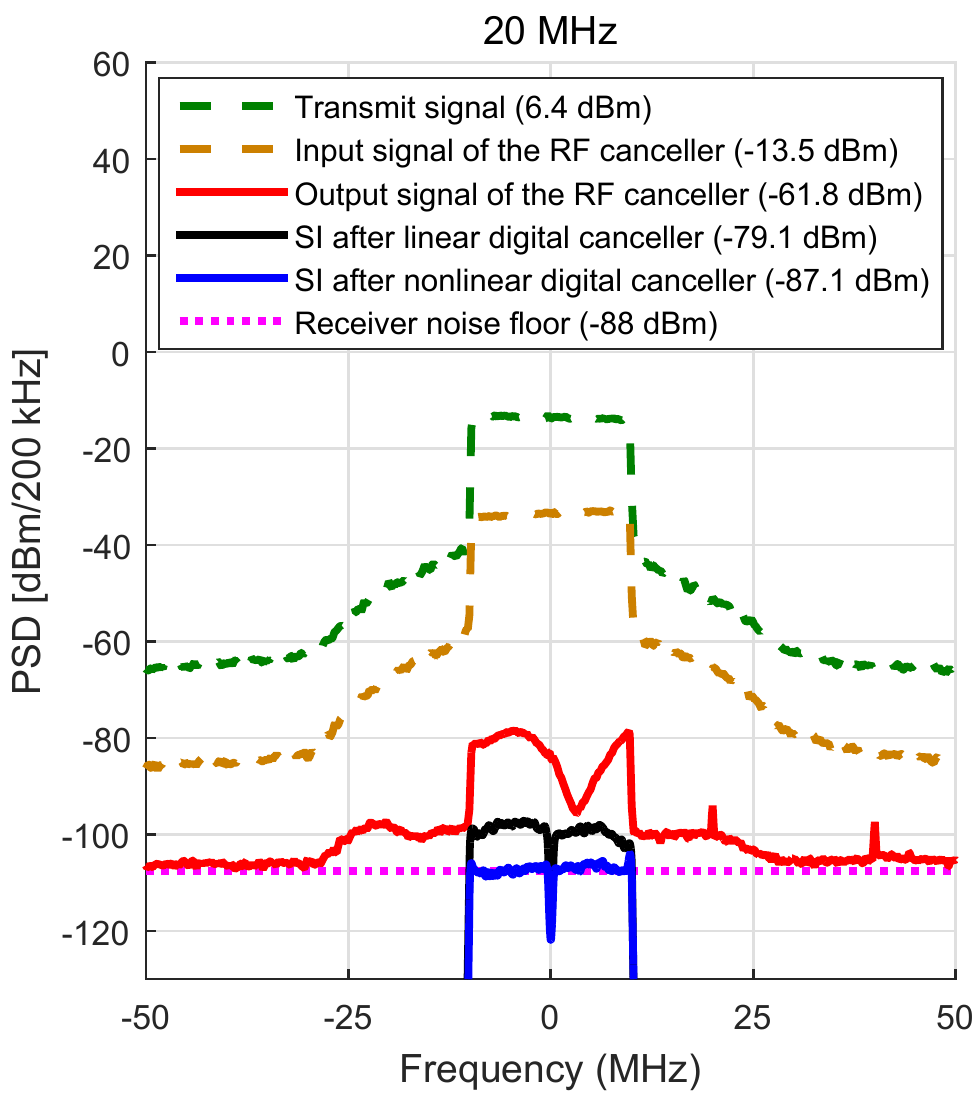}	
								\caption{}
                \label{fig:total_cancel_20M}
        \end{subfigure}%
        ~
        \begin{subfigure}[t]{0.32\textwidth}
                \includegraphics[width=\textwidth]{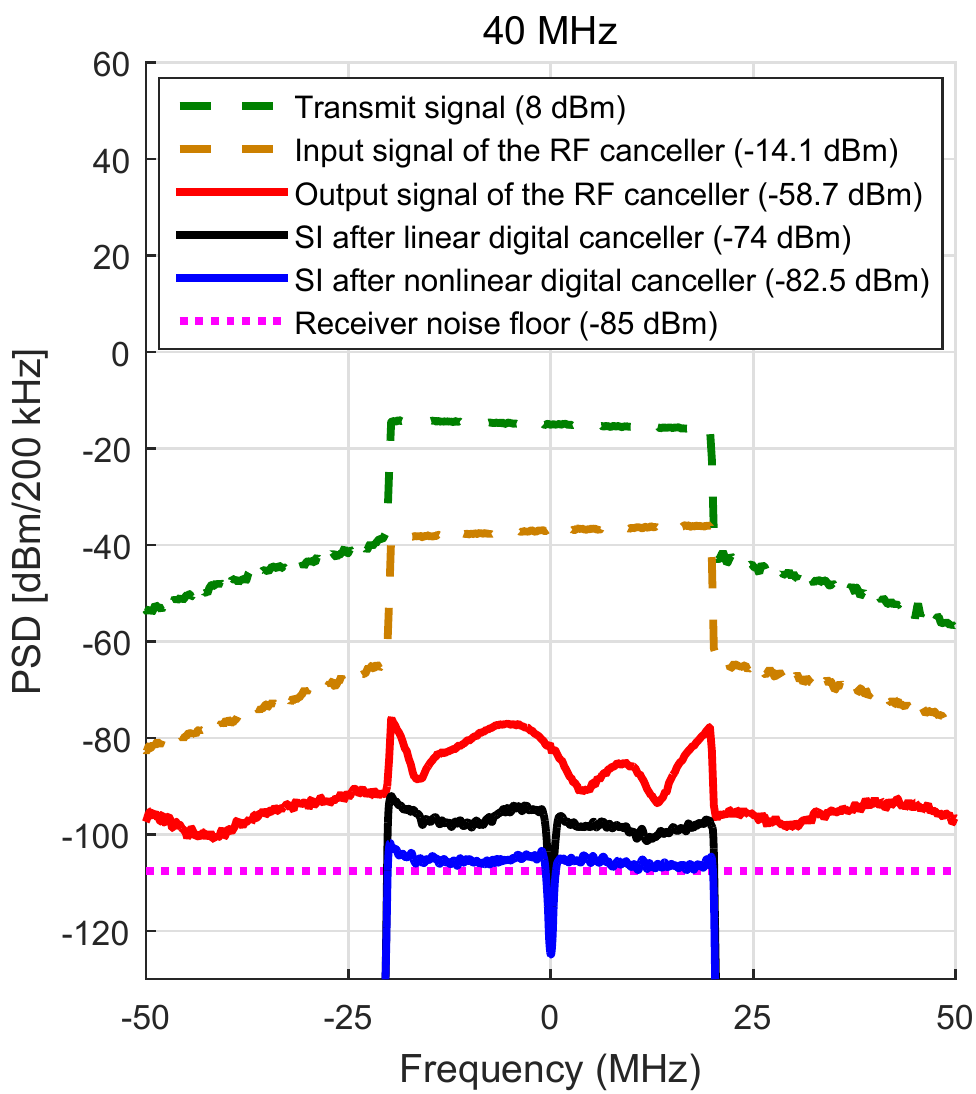}
								\caption{}
                \label{fig:total_cancel_40M}
        \end{subfigure}
				~
        \begin{subfigure}[t]{0.32\textwidth}
                \includegraphics[width=\textwidth]{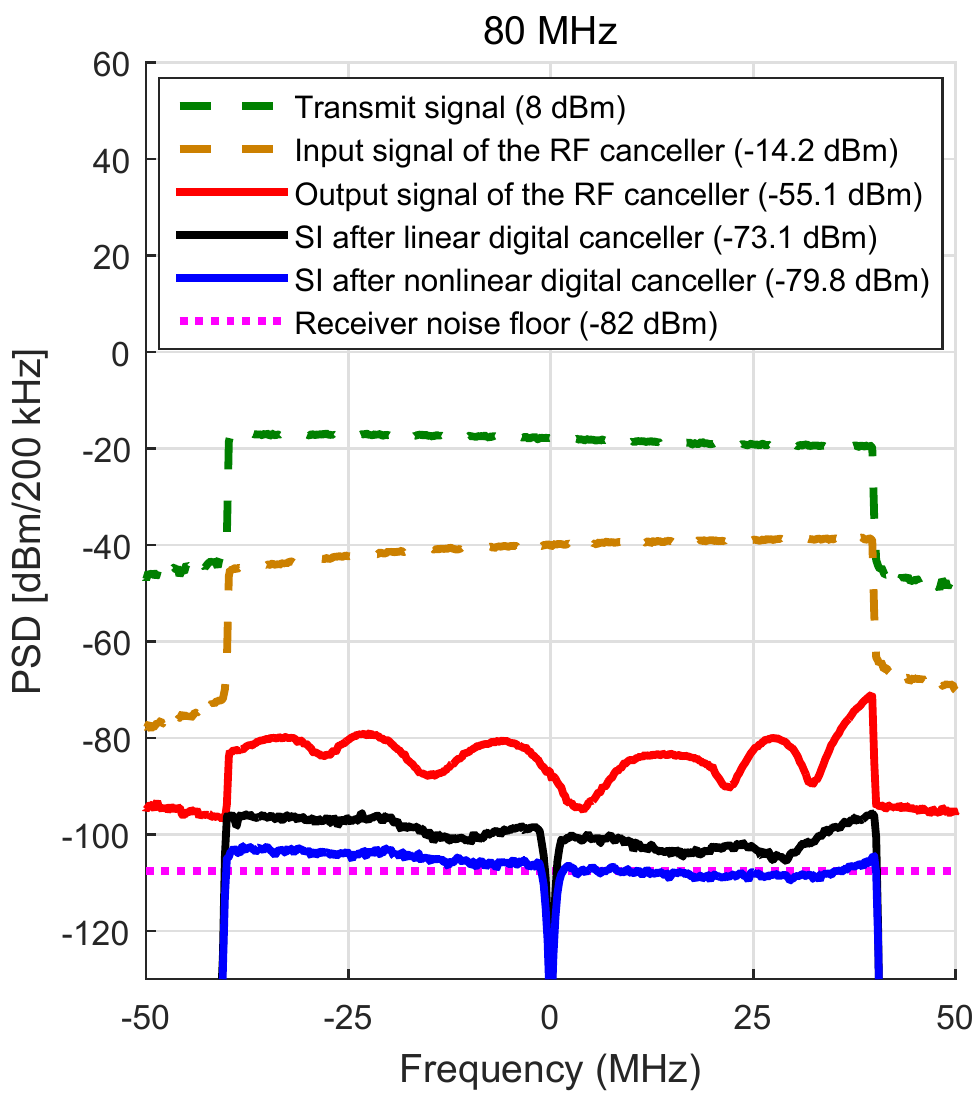}
								\caption{}
                \label{fig:total_cancel_80M}
        \end{subfigure}
        \caption{{\revColor{}The overall cancellation performance with (a) 20~MHz, (b) 40~MHz,  and (c) 80~MHz LTE waveforms at 2.46~GHz, the contribution of each cancellation stage being shown separately (projected on to the baseband).} In the digital canceller, nonlinearities up to the 11th order are considered to accurately model the low-cost power amplifier. This allows the nonlinear digital canceller to push the residual SI to the receiver noise floor, clearly outperforming the linear canceller whose performance is heavily limited by the power amplifier induced nonlinear distortion.}
\label{fig:total_canc}
\end{figure*}


{\revColor{}Despite the impressive RF cancellation performance demonstrated above, the level of the residual SI is still substantially above the receiver noise floor in the digital domain. Thus, the downconverted and digitized received signal is next processed by the digital SI canceller, which is implemented in a host processor. Furthermore, because of the low-cost mobile-scale PA, nonlinear digital SI cancellation with LMS parameter learning is utilized \cite{Korpi15}.

Figures~\ref{fig:total_canc}(\subref{fig:total_cancel_20M}),~\ref{fig:total_canc}(\subref{fig:total_cancel_40M}), and~\ref{fig:total_canc}(\subref{fig:total_cancel_80M}) show example signal spectra of the SI signal after each cancellation stage, including adaptive nonlinear digital cancellation described in Section~\ref{sec:nl_canc}. It can be observed that, for the considered bandwidths of 20,~40, and 80~MHz, the combined attenuation of the circulator and the RF canceller is 63--68~dB, after which the described adaptive nonlinear digital canceller attenuates the SI signal further by another 25~dB. With classical linear digital cancellation, the total SI attenuation is roughly 10~dB less, illustrating that nonlinear digital cancellation processing is a necessary requirement for a mobile full-duplex transceiver with substantial nonlinear distortion in the transmitter power amplification stage. 

Overall, these examples demonstrate that the described mobile full-duplex transceiver architecture is capable of attenuating the SI practically to the receiver noise floor even with the 80~MHz signal bandwidth. This indicates that very good overall wideband performance is achieved also under practical conditions, despite a heavily nonlinear transmitter PA and a shared antenna. Hence, the proposed architecture is well capable of coping with the challenges posed by the mobile device environment, such as circuit imperfections, wideband operation, and time-varying channel conditions.}

\section{Conclusion}
\label{sec:conc}

To fully capitalize the benefits of inband full-duplex radio technology in, e.g., cellular networks, also the mobile devices should support simultaneous transmission and reception at the same center-frequency. This article explored the most prominent challenges in implementing mobile inband full-duplex devices, and also described solutions to these problems. In particular, a mobile full-duplex device must be capable of shared-antenna operation, as well as adapting to the changes in the channel environment. Also, since a mobile-scale device is typically relying on low-cost components, the different circuit impairments must be considered as they directly affect the self-interference cancellation ability. Furthermore, nowadays also mobile devices must be able to handle very wideband signals to ensure high data rates.

As a solution to these challenges, this article demonstrated a prototype implementation having a shared transmit/receive antenna, an adaptive wideband multi-tap RF canceller, and an adaptive wideband nonlinear digital canceller. It was shown that the novel wideband RF cancellation solution with built-in capability to automatically track changes in the self-interference channel characteristics can yield RF cancellation gains beyond 40~dB, even with waveform bandwidths in the order of 80~MHz and a highly nonlinear low-cost power amplifier. Measured examples also demonstrated good self-adaptation capabilities against fast changes in the environment around the antenna. Moreover, the adaptive nonlinear digital canceller was shown to push the residual self-interference down to the noise floor of the receiver, also under a heavily nonlinear transmitter power amplifier. All in all, these findings pave the way towards potentially enabling the full-duplex capability also in the mobile devices of the future 5G or beyond radio communication systems.

\bibliographystyle{IEEEtran}
\bibliography{IEEEabrv,IEEEref}

\begin{thebibliography}{10}
\providecommand{\url}[1]{#1}
\csname url@samestyle\endcsname
\providecommand{\newblock}{\relax}
\providecommand{\bibinfo}[2]{#2}
\providecommand{\BIBentrySTDinterwordspacing}{\spaceskip=0pt\relax}
\providecommand{\BIBentryALTinterwordstretchfactor}{4}
\providecommand{\BIBentryALTinterwordspacing}{\spaceskip=\fontdimen2\font plus
\BIBentryALTinterwordstretchfactor\fontdimen3\font minus
  \fontdimen4\font\relax}
\providecommand{\BIBforeignlanguage}[2]{{%
\expandafter\ifx\csname l@#1\endcsname\relax
\typeout{** WARNING: IEEEtran.bst: No hyphenation pattern has been}%
\typeout{** loaded for the language `#1'. Using the pattern for}%
\typeout{** the default language instead.}%
\else
\language=\csname l@#1\endcsname
\fi
#2}}
\providecommand{\BIBdecl}{\relax}
\BIBdecl

\bibitem{Sabharwal14}
A.~Sabharwal, P.~Schniter, D.~Guo, D.~Bliss, S.~Rangarajan, and R.~Wichman,
  ``In-band full-duplex wireless: Challenges and opportunities,'' \emph{IEEE
  Journal on Selected Areas in Communications}, vol.~32, no.~10, Oct. 2014.

\bibitem{Duarte12}
M.~Duarte, C.~Dick, and A.~Sabharwal, ``Experiment-driven characterization of
  full-duplex wireless systems,'' \emph{IEEE Transactions on Wireless
  Communications}, vol.~11, no.~12, pp. 4296--4307, Dec. 2012.

\bibitem{Goyal15a}
S.~Goyal, P.~Liu, S.~S. Panwar, R.~A. Difazio, R.~Yang, and E.~Bala, ``Full
  duplex cellular systems: will doubling interference prevent doubling
  capacity?'' \emph{IEEE Communications Magazine}, vol.~53, no.~5, pp.
  121--127, May 2015.

\bibitem{Andrews14}
J.~Andrews, S.~Buzzi, W.~Choi, S.~Hanly, A.~Lozano, A.~Soong, and J.~Zhang,
  ``What will {5G} be?'' \emph{IEEE Journal on Selected Areas in
  Communications}, vol.~32, no.~6, pp. 1065--1082, Jun. 2014.

\bibitem{Hong14}
S.~Hong, J.~Brand, J.~Choi, M.~Jain, J.~Mehlman, S.~Katti, and P.~Levis,
  ``Applications of self-interference cancellation in {5G} and beyond,''
  \emph{IEEE Communications Magazine}, vol.~52, no.~2, pp. 114--121, Feb. 2014.

\bibitem{Bharadia13}
D.~Bharadia, E.~McMilin, and S.~Katti, ``Full duplex radios,'' in \emph{Proc.
  SIGCOMM'13}, Aug. 2013, pp. 375--386.

\bibitem{Everett11}
E.~Everett, M.~Duarte, C.~Dick, and A.~Sabharwal, ``Empowering full-duplex
  wireless communication by exploiting directional diversity,'' in \emph{Proc.
  45th Asilomar Conference on Signals, Systems and Computers}, Nov. 2011, pp.
  2002--2006.

\bibitem{Sahai13}
A.~Sahai, S.~Diggavi, and A.~Sabharwal, ``On degrees-of-freedom of full-duplex
  uplink/downlink channel,'' in \emph{IEEE Information Theory Workshop (ITW)},
  Sep. 2013, pp. 1--5.

\bibitem{Korpi15}
D.~Korpi, Y.-S. Choi, T.~Huusari, S.~Anttila, L.~Talwar, and M.~Valkama,
  ``Adaptive nonlinear digital self-interference cancellation for mobile inband
  full-duplex radio: Algorithms and {RF} measurements,'' in \emph{Proc. IEEE
  Global Communications Conference (GLOBECOM)}, Dec. 2015.

\bibitem{Choi13}
Y.-S. Choi and H.~Shirani-Mehr, ``Simultaneous transmission and reception:
  Algorithm, design and system level performance,'' \emph{IEEE Transactions on
  Wireless Communications}, vol.~12, no.~12, pp. 5992--6010, Dec. 2013.

\bibitem{Korpi13}
D.~Korpi, T.~Riihonen, V.~Syrj\"{a}l\"{a}, L.~Anttila, M.~Valkama, and
  R.~Wichman, ``Full-duplex transceiver system calculations: Analysis of {ADC}
  and linearity challenges,'' \emph{IEEE Transactions on Wireless
  Communications}, vol.~13, no.~7, pp. 3821--3836, Jul. 2014.

\bibitem{Korpi133}
D.~Korpi, L.~Anttila, V.~Syrj\"{a}l\"{a}, and M.~Valkama, ``Widely linear
  digital self-interference cancellation in direct-conversion full-duplex
  transceiver,'' \emph{IEEE Journal on Selected Areas in Communications},
  vol.~32, no.~9, pp. 1674--1687, Sep. 2014.

\bibitem{Debaillie14}
B.~Debaillie, D.-J. van~den Broek, C.~Lavin, B.~van Liempd, E.~Klumperink,
  C.~Palacios, J.~Craninckx, B.~Nauta, and A.~Parssinen, ``Analog/{RF}
  solutions enabling compact full-duplex radios,'' \emph{IEEE Journal on
  Selected Areas in Communications}, vol.~32, no.~9, pp. 1662--1673, Sep. 2014.

\bibitem{Isaksson06}
M.~Isaksson, D.~Wisell, and D.~Ronnow, ``A comparative analysis of behavioral
  models for {RF} power amplifiers,'' \emph{IEEE Transactions on Microwave
  Theory and Techniques}, vol.~54, no.~1, pp. 348--359, Jan. 2006.

\bibitem{Syrjala13}
V.~Syrj\"{a}l\"{a}, M.~Valkama, L.~Anttila, T.~Riihonen, and D.~Korpi,
  ``Analysis of oscillator phase-noise effects on self-interference
  cancellation in full-duplex {OFDM} radio transceivers,'' \emph{IEEE
  Transactions on Wireless Communications}, vol.~13, no.~6, pp. 2977--2990,
  Jun. 2014.

\end{thebibliography}

\end{document}